\documentstyle[12pt,epsfig,amsmath,amssymb]{article}
\begin{document}

\title{\bf Decompositional equivalence}

\author{{Chris Fields}\\ \\
{\it 21 Rue des Lavandi\`eres}\\
{\it Caunes Minervois 11160 France}\\ \\
{fieldsres@gmail.com}}
\maketitle

\begin{abstract}
Both classical and quantum mechanics assume that physical laws are invariant under changes in the way that the world is labeled.  This \textit{Principle of Decompositional Equivalence} is formalized, and shown to forbid finite experimental demonstrations that given pointer states $|p_{1}>$ and $|p_{2}>$ refer to the same physical system $S$.  It is then shown that any extra-theoretical assumption that given pointer states $|p_{1}>$ and $|p_{2}>$ with indistinguishable coefficients in a Schmidt basis for the universe refer to the same physical system $S$ is stronger than the assumption of classicality.  Standard interpretations of quantum mechanics make such as assumption in analyzing measurement; hence they are logically inconsistent.
\end{abstract}

\textbf{Keywords} Foundations of quantum mechanics, measurement problem, existential interpretation, quantum Darwinism

\section{Introduction}

Both classical and quantum mechanics assume that physical laws are invariant under changes in the way that the world is labeled.  Let us call this assumption the ``Principle of Decompositional Equivalence (PDE)."   It can be formalized as follows.  Consider a universe $U$, and let $S^{0} \subset U$ be a system comprising a number of quantum degrees of freedom that is larger than one, but negligibly small compared to the degrees of freedom of its environment $E^{0} = U \setminus S^{0}$ (here and below, ``$X = Z \setminus Y$" is used as a shorthand indicating $Z = X \otimes Y$).  Consider a family $\mathcal{F}^{\mathit{0}}$ of distinct systems such that each $S^{k} \in \mathcal{F}^{\mathit{0}}$ can be written $S^{k} = S^{0} \otimes \psi^{k}$, where $\psi^{k}$ represents a single degree of freedom not part of $S^{0}$.  Corresponding to each $S^{k}$ is an environment $E^{k} = U \setminus S^{k}$.  In this notation, the PDE is:

\begin{quote}
\textit{Principle of Decompositional Equivalence}: For any system $S^{0} \subset U$ and any family of systems $\mathcal{F}^{\mathit{0}}$ as defined as above, if $\mathcal{H}_{\mathit{S^{0}-E^{0}}} \neq \mathrm{0}$ is a well-defined Hamiltonian specifying an interaction between $S^{0}$ and $E^{0}$, there are well-defined Hamiltonians $\mathcal{H}_{\mathit{S^{k}-E^{k}}} \neq \mathrm{0}$ specifying interactions between $S^{k}$ and $E^{k}$ for every $S^{k} \in \mathcal{F}^{\mathit{0}}$.
\end{quote}

Fully unitary quantum mechanics clearly requires a universe compliant with the PDE.  Without it, neither the Hamiltonian $\mathcal{H}_{\mathit{U}}$ of the universe $U$ as a whole, nor any observables expressible in terms of $\mathcal{H}_{\mathit{U}}$ are well-defined.

The PDE states that decompositional boundaries are not respected by physical dynamics, i.e. that they are arbitrary.  It therefore conflicts \textit{prima facie} with the intuitive, classical notion that the physical world includes discrete ``objects" with real, ``natural" boundaries.  It also conflicts \textit{prima facie} with the quantum mechanical notion that decoherence acting at the ``natural" boundaries of macroscopic objects einselects pointer states that then define those ``natural" boundaries for observers, as has been proposed under the rubrics of the existential interpretation and quantum Darwinism \cite{zurek98rough, zurek03rev, zurek09rev, schloss}.  This paper shows that this \textit{prima facie} conflict is serious.  It demonstrates two limitations imposed on observers by the PDE: first, that no observer can conclusively identify the system $S$ represented by a given pointer state $|p>$, even if the non-zero components of $|p>$ are expressed in a Schmidt basis for $U$; and second, that no two observers at one time, or one observer at two times, can conclusively establish that two pointer states $|p_{1}>$ and $|p_{2}>$ represent the same system $S$, even if their non-zero components in a Schmidt basis for $U$ are identical.  Hence any interactions between observers and macroscopic ``objects" that involve either re-identification of objects by observers over time, or agreement among observers that they are interacting with the same object, require an extra-theoretical assumption that \textit{complete} specifications of pointer states are ``given" by decoherence, or by any other process that acts effectively as an oracle.  This extra-theoretical assumption of oracular completeness is characterized, and shown to be stronger than the assumption of classicality.  All interpretations of quantum mechanics in which the pointer states of macroscopic objects such as apparatus are treated as ``given" by an oracular process, and in which the measurement problem thereby arises in its traditional form, assume oracular completeness, and are therefore logically inconsistent.  Dropping the assumption of oracular completeness effectively converts the measurement problem into a conceptually more straightforward problem of understanding how information is encoded by the states of observers.

\section{Decoherence as oracle: ``frog perspective"}

The notion of measurement, indeed the notion of an independently-existing external world, requires that not all facts are stipulated; some facts must be ``given" by observation.  Such ``given" facts are, in effect, provided by an oracle.  In the context of minimal quantum mechanics, decoherence \cite{zurek98rough, zurek03rev, schloss} is the oracle.  We first consider the functioning of the oracle of decoherence from the point of view of a fixed observer $O$, what Tegmark \cite{Tegmark10} calls the ``frog perspective."  Suppose $O$ makes a sequence of $N$ observations, each of finite duration $\Delta t$.  The information contributing to an observation initiated at time $t_{i}$ has a maximum transmission channel pathlength of $c \Delta t$, where $c$ is the speed of light.  All such information, regardless of its ultimate origin, must therefore be encoded by degrees of freedom within a sphere of radius $c \Delta t$ surrounding the fixed position of $O$ at $t_{i}$ if it is to contribute to the observation initiated at $t_{i}$.  Call the collection of degrees of freedom that contribute information to $O$'s observation at $t_{i}$ $Z^{O}|_{t_{i}}$, the ``horizon" of $O$ at $t_{i}$.  For any $\Delta t$ small compared to the age of the universe but larger than about $1 \, ps$, $Z^{O}|_{t_{i}}$ is macroscopic, but still small compared to its environment $U \setminus Z^{O}|_{t_{i}}$.  Hence provided that $Z^{O}|_{t_{i}}$ is not isolated, the interaction between $Z^{O}|_{t_{i}}$ and $U \setminus Z^{O}|_{t_{i}}$ will einselect a pointer state of $Z^{O}|_{t_{i}}$ within a decoherence interval of at most the photon-transit time $\Delta t$.  Call this pointer state $|p_{i}>$.  The observations ``given" to any $O$ by the oracle of decoherence constitute a sequence of such pointer states $|p_{i}>$.

We now ask, what can $O$ infer from these ``given" observations $|p_{1}> ... |p_{N}>$, and in particular, what can $O$ infer about the ``object" $Z^{O}$ that those observations describe, an object that properly includes $O$ herself?  We assume fully unitary quantum mechanics, and hence the PDE.  Employing the notation used to formalize the PDE, two lemmas follow immediately from this assumption:

\textit{Lemma 1}: Within any subset $U^{\prime} \subseteq U$ with finite degrees of freedom, if at least one macroscopic system is not isolated, no macroscopic systems are isolated.  \textit{Proof}: Let $S^{0}$ name the non-isolated system.  Then by the PDE, each element of $\mathcal{F}^{\mathit{0}}$ is non-isolated.  This labeling process can be iterated to incorporate every degree of freedom of $U^{\prime}$ into the family $\mathcal{F}^{\mathit{m}}$ associated with some non-isolated $S^{m}$. $\square$

\textit{Lemma 2}: If $\mathcal{H}_{\mathit{S^{0}-E^{0}}}$ decoheres $S^{0}$ and einselects pointer states $\{|p^{0}_{i}>\}$, then for every $S^{k} \in \mathcal{F}^{\mathit{0}}$, $\mathcal{H}_{\mathit{S^{k}-E^{k}}}$ decoheres $S^{k}$ and einselects pointer states $\{|p^{k}_{i}>\}$.  \textit{Proof}: Let $|S^{0}>$ be an arbitrary state of $S^{0}$, and $|\psi^{k}>$ be an arbitrary state of a degree of freedom $\psi^{k}$ not contained within $S^{0}$.  Because $<S^{0}|\psi^{k}> = 0$, $\mathcal{H}_{\mathit{S^{k}-E^{k}}}$ can be defined such that $|S^{0}><S^{0}|\mathcal{H}_{\mathit{S^{k}-E^{k}}} = \mathcal{H}_{\mathit{S^{0}-E^{0}}}$, in which case the required pointer states are $|p^{k}_{i}> = (1 - |\psi^{k}|)^{2} |p^{0}_{i}> + \mathcal{H}_{\mathit{S^{k}-E^{k}}} |\psi^{\mathit{k}}>$.  Decoherence of $S^{k}$ and hence einselection of some $|p^{k}_{i}>$ depends only on $E^{k}$ being sufficiently large and $\mathcal{H}_{\mathit{S^{k}-E^{k}}}$ being non-zero.  $E^{0}$ is sufficiently large by assumption, and $E^{k}$ is only one degree of freedom smaller.  If $\mathcal{H}_{\mathit{S^{0}-E^{0}}}$ is non-zero, $\mathcal{H}_{\mathit{S^{k}-E^{k}}}$ is non-zero by Lemma 1. $\square$

Lemma 1 shows that if any $Z^{O}|_{t_{i}}$, is isolated, all are isolated, and in fact all are isolated for all observers.  Hence the only non-trivial case is the one in which no $Z^{O}|_{t_{i}}$ for any observer is isolated.  In this case, all observers $O$ are provided by the oracle of decoherence with pointer states $|p^{O}_{i}>$ for every $\Delta t$-long interval of observation.  Because the $|p^{O}_{i}>$ are states of non-isolated systems, the no-cloning theorem \cite{no-clone} does not forbid their cloning and hence re-identification by observers.  The ``given" $|p^{O}_{i}>$ are not, however, einselected at a stipulated $S-O$ boundary by a characterized $S-O$ interaction; they are einselected at the boundary between $Z^{0}|_{t_{i}}$ and $U \setminus Z^{0}|_{t_{i}}$, a boundary to which $O$ has, by definition, no independent observational access.  Hence any observer $O$ presented with a $|p_{j}>$ identical to a previously-observed $|p_{i}>$ is faced with the question of whether $|p_{j}>$ and $|p_{i}>$ refer to the same system, i.e. the question of whether $Z^{0}|_{t_{i}} = Z^{0}|_{t_{j}}$.  This question is not trivial.  Any observer walking into a laboratory, for example, will want to know whether pointer positions on meters and numbers displayed on digital readouts represent experimental results or calibration tests \cite{qdar}.  Whether this question can be answered definitively depends on what kind of oracle decoherence is.

Assume that $\mathcal{H}_{U}$ is known completely, and consider two distinct types of oracle.  A ``complete" oracle, when queried at time $t_{i}$ by an observer $O$, provides the complete specification of a pointer state $|p_{i}>$ in a basis for $Z^{O}|_{t_{i}}$ that is a projection into $Z^{O}|_{t_{i}}$ of a Schmidt basis for $U$, including all components with coefficients of zero.  In contrast, a ``projecting" oracle, when queried at time $t_{i}$, provides only the non-zero Schmidt-basis components of the same pointer state.   An observer $O$ who knowingly queries a complete oracle at successive times $t_{1} ... t_{N}$ and receives pointer-state specifications $|p^{C}_{1}> ... |P^{C}_{N}>$ can infer, from the set of basis vectors appearing in the specification of each $|p^{C}_{i}>$, what system $S^{i}$ each $|p^{C}_{i}>$ represents, and can unambiguously identify pairs of pointer states that represent the same system.  However, if $O$ knowingly queries a projecting oracle at successive times $t_{1} ... t_{N}$ and receives pointer-state specifications $|p^{P}_{1}> ... |P^{P}_{N}>$, $O$ cannot infer the existence of any system unambiguously, and cannot identify pairs of (projected) pointer states that represent the same system, even if their components are identical.

\textit{Theorem 1}: No oracle $\mathcal{R}$ can be demonstrated to be complete by a finite number of experiments.  \textit{Proof}: Let $|p_{1}> ... |p_{N}>$ be the pointer states ``given" by $\mathcal{R}$ to $O$ in response to a finite number of queries.  (1) Because $\mathcal{R}$ must answer each query in finite time $\Delta t$, $\mathcal{R}$ can only specify each $|p_{i}>$ using a finite number of basis vectors with coefficients given to some finite accuracy $\delta$.  Hence $O$ cannot prove that the set of components comprising a ``given" pointer state $|p_{i}>$ is complete by a normalization test.  (2) Since each $|p_{i}>$ has a finite specification, $O$ can compute the projection $|p_{1}><p_{1}|p_{i}>$ for each given $|p_{i}>$.  If for any query $i$, $|p_{1}><p_{1}|p_{i}> \neq |p_{i}>$, $\mathcal{R}$ is a projecting oracle.  Hence the only case of interest is that in which $|p_{1}><p_{1}|p_{i}> = |p_{i}>$ for all $i$, that is, the case in which the basis of, and hence the boundary of $Z^{0}$ appears static.  (3) The PDE requires that for any system $S^{0}$, a family $\mathcal{F}^{\mathit{0}}$ of systems $S^{k}$ differing from $S^{0}$ by only one degree of freedom $\psi^{k}$ can be defined; Lemma 2 requires that any such system $S^{k}$ is represented by pointer states.  Demonstrating that $\mathcal{R}$ is complete for an apparently static horizon $Z^{0}$ by direct empirical observation therefore requires demonstrating that any degree of freedom $\psi^{k}$ not represented by a basis vector in $|p_{1}>$ is outside of $Z^{0}$, i.e. that the minimum information transfer pathlength from $\psi^{k}$ to $O$ is greater than $c \Delta t$.  $O$ cannot do this, however, as $O$ has no independent observational access to the boundary of $Z^{0}|_{t_{i}}$ for any $t_{i}$.  (4) Direct observation of the physical process by which $\mathcal{R}$ generates the $|p_{i}>$ thus being impossible, $\mathcal{R}$ can only be treated as an algorithm to be reverse engineered.  If $\mathcal{R}$ can be shown to implement no algorithmic step in which a component $\alpha_{ij} u_{ij}$ of a pointer state $|p_{i}>$ is tested to determine whether $|\alpha_{ij}| \leq \delta$, $\mathcal{R}$ can be shown to be complete.  However, any $\mathcal{R}$ capable of implementing such a step has at least the computational complexity of a classical finite state machine.  Any reverse-engineering demonstration of $\mathcal{R}$'s algorithm is, therefore, forbidden by Theorem 2 of Moore \cite{moore56}, which states that no finite set of observations can unambiguously determine the algorithm executed by a finite state machine. $\square$

Theorem 1 applies without modification not only to the full pointer state $|p_{i}>$ of an observer's horizon $Z^{O}|_{t_{i}}$ at $t_{i}$, but to any projection $|S><S|p_{i}>$ that implicitly defines a system $S \subset Z^{O}|_{t_{i}}$ that an observer might stipulate to be ``of interest" at $t_{i}$.  Theorem 1 therefore shows that the situation envisaged by quantum Darwinism, in which a collection of independent observers interact with ``given" substates of their own individual, disjoint horizons in order to mutually and non-perturbatively establish the state of an external system of interest \cite{zurek03rev, zurek09rev}, cannot be realized without an extra-theoretical and non-empirical assumption that the oracle of decoherence is complete.  The same restriction applies to a single observer who compares a currently-observed pointer state to a memory record of a previously-observed pointer state in order to establish the continuing existence of an identified $S$, as envisged by the existential interpretation \cite{zurek98rough, zurek03rev, schloss}; any such comparison of ``given" pointer states across measurement times requires an extra-theoretical, non-empirical assumption of oracular completeness.  Nothing in the proof of Theorem 1, moreover, relies on unitary evolution of $U$ as a whole, or on the specific mechanism of decoherence; hence Theorem 1 applies to any postulated oracular mechanism that delivers pointer-state specifications in a fixed basis and in a fixed amount of time, provided only that such a mechanism can be represented as an observable in a universe compliant with the PDE.

\section{Decoherence as oracle: ``bird perspective"}

The restrictions on observation implied by Theorem 1 can also be reached from Tegmark's meta-theoretical ``bird perspective" by noting that a simple notational change:

\begin{equation}
\Psi = \sum \lambda_{k} |s_{k}> \quad \longrightarrow \quad \Psi = \sum \lambda_{k} |S^{k}>
\end{equation}

converts a wave function specifed as a superposition of basis vectors spanning a stipulated system $S$ into a wave function specified as a superposition of individual states $|S^{k}>$ of a collection $\{S^{k} \subset S\}$ of stipulated subsystems such that $\forall k, dim(S^{k}) = dim(S) - 1$ and $\cup\{S^{k}\} = S$.  Let $\mathcal{O}$ be an observable defined everywhere in a $U \supset S$ compliant with decompositional equivalence, and for each $k$, suppose $S^{k}$ has a pointer state $|p^{k}>$ with an associated eigenvalue $\alpha^{k}$ such that $\mathcal{O} \mathit{|p^{k}> = \alpha^{k} |p^{k}>}$.  Then 

\begin{equation}
\mathcal{O} \mathit{\Psi} = \sum \lambda_{k} \alpha^{k} |p^{k}>.
\end{equation}

Now suppose there is a $|p^{0}> \in \{|p^{k}>\}$ such that $\forall k, |\alpha^{0} - \alpha^{k}| \leq \delta$ and $<p^{0}|p^{k}> \geq (1 - \delta)$, i.e. the $\alpha^{k} |p^{k}>$ are indistinguishable up to a criterion $\delta$.  In this case,

\begin{equation}
\mathcal{O} \mathit{\Psi} \sim \mathcal{O} \mathit{|S^{0}>}
\end{equation}

up to the uncertainty $\delta$, i.e. an observer $O$ ``given" $|p^{0}>$ with uncertainty $\delta$ by $\mathcal{O}$ considered as an oracle cannot determine whether $\mathcal{O}$ has acted on a single system $S^{0}$ or on an entangled family of systems $\sum \lambda_{k} |S^{k}>$.

The above notion of a superposition of systems is clearly just the system-side dual of the idea that observables may be represented as POVMs; here the $\mathcal{O^{\mathit{k}}} = \mathit{|S^{k}><S^{k}|\mathcal{O}}$ are the POVM components.  Theorem 1 states, in this notation, that an observer employing an observable $\mathcal{O}$ to characterize a system $S$ cannot determine by observation which component $\mathcal{O^{\mathit{k}}}$ generated the pointer state $|p_{j}>$ of $S$ ``given" by the $j^{th}$ instance of the action of $\mathcal{O}$.

The notational change illustrated in Eq. 1 suggests a non-standard interpretation of decoherence.  Decoherence is standardly thought of as dissipating quantum coherence into the environment \cite{zurek98rough, zurek03rev, schloss}.  The question is, the environment of what?  If an observer $O$ stipulates a system $S$ and hence a system - environment boundary, decoherence acts as an oracle to ``give" $O$ a pointer state $|p>$ for $S$.  But like all gifts from oracles, this one has a cost.  The cost is that $O$ cannot distinguish $S$ from a superposition of all possible systems with pointer states indistinguishable, at $O$'s available resolution, from $|p>$.  The boundary of $S$ has lost the precision of stipulation, and taken on the quantum fuzziness of observation.  Thus even from $O$'s perspective, decoherence has not destroyed quantum coherence, nor has it moved it into some distant ``environment"; it has just transferred quantum coherence from $S$'s \textit{state} to $S$'s \textit{boundary}.  The assumption of oracular completeness for decoherence is, effectively, the assumption that $|p>$ specifies $S$'s boundary as fixed, not fuzzy.  But if decoherence merely moves coherence from states to boundaries, this would imply that $S$'s state had no coherence to begin with, i.e. that $S$ was not a quantum system at all, but rather was classical.

\section{Oracular completeness implies classicality}

Although Theorem 1 shows that oracular completeness cannot be experimentally demonstrated by $O$, it may be supposed that it is harmless as an assumption.  As suggested by the reasoning above, this is not the case.  

\textit{Theorem 2}: If a universe $U$ includes a complete oracle $\mathcal{R}$ for at least two non-commuting observables, $U$ is classical.  \textit{Proof}:  Assume a pointer state $|p>$ for an observable $\mathcal{O}$ acting on a system $S$ is given in a Schmidt basis $\{u_{i}\}$ for $U$ as $|p> = \sum_{k=1}^{n} \alpha_{k} u_{k}$.  The PDE requires that for any such system $S$, there are an arbitrary number of alternative systems $S^{k}$ such that $|S^{k}> = |S>|\psi^{k}>$ and $<\psi^{k}|S> = 0$.  For any such $S^{k}$, and for any uncertainty criterion $\delta$, there is a pointer state $|p^{k}>$ of $S^{k}$ such that $|p^{k}> = (1 - \epsilon)^{2} |p> + \, \epsilon u_{m}$ for some $m \geq n$ and $\epsilon \leq \delta$.  To guarantee that $|p>$ is distinguishable from $|p^{k}>$ when given by $\mathcal{R}$, it can only be the case that the amplitude $\epsilon = 0$.  But if the uncertainites of physical state specifications generated by non-commuting observables are identically zero, or even guaranteed to be less than any required $\delta$, $U$ can only be classical.  $\square$

The assumption of oracular completeness is, however, even stronger than an assumption of classicality.  The formalism of classical physics is consistent with and in fact relies on the PDE, as evidenced in practice by the ability to consider and label arbitrary volumes of a classical fluid or regions of a classical phase space.  Moreover, nothing in the formalism of classical physics requires oracular completeness.  Classical physics is consistent with arbitrary ambiguities in the relationship between ``given" states of pointers and actual states of the world, as generations of experimentalists have experienced to their chagrin.  Oracular completeness is an extra-theoretical, non-empirical assumption even in classical physics.  This is not surprising: oracular completeness is, at bottom, an assumption about the \textit{knowledge} that observers obtain from the oracle of ``given" observations.  Theorem 2 states that certainty about all physical states can only be obtained in a classical world.  But a classical world hardly implies certainty.

\section{Oracular completeness in the measurement problem}

We now turn to the role that implicit assumptions of oracular completeness have played, and continue to play, in the project of interpreting quantum mechanics.  Consider Schr\"odinger's unfortunate cat.  Two observations are made, one when the cat is put into the box, and one when the box is subsequently opened.  The cat is in a pointer state, either $|dead>$ or $|alive>$, during each observation.  While the box is closed, the cat is assumed on the basis of a theoretical calculation to exist in a superposition of $|dead>$ and $|alive>$.  Schr\"odinger's cat, like everything else, is a quantum system; it is the oracle of decoherence that resolves it into one of its two pointer states whenever the box is open.

This account of the \textit{gedankenexperiment} with Schr\"odinger's cat assumes oracular completeness.  One and the same cat goes into the box, remains in the box when the cover is closed, and is observed, in one or the other of its pointer states, when the box is re-opened.  The fixed identity of the cat, not just its state, is taken to be given by observation.  What justifies this assumption?  It is not required by the formalism of quantum mechanics.  It is not proved by observation; Theorem 1 forbids such proof given pointer states, and no observations are attempted while the box is closed.  The assumption that one and the same cat is present throughout the experiment is based on ``common sense" - it is extra-theoretical and non-empirical.  It is also, as Theorem 2 shows, flatly inconsistent with the assumption that Schr\"odinger's cat is a quantum system.

An implicit assumption of oracular completeness when describing measurement is not specific to Schr\"odinger's cat.  Writing down a von Neumann chain:

\begin{align}
(\sum_{i}\lambda_{i}|s_{i}>)|A^{ready}>|E_{init}>|O^{ready}>\,&\rightarrow \sum_{i}\lambda_{i}|s_{i}>|a_{i}>|e_{i}>|o_{i}> \nonumber \\
&\rightarrow |s_{n}>|a_{n}>|e_{n}>|o_{n}>
\end{align}

explicitly assumes that the system of interest $S = \sum_{i}\lambda_{i}|s_{i}>$, the apparatus $A$, observer $O$ and environment $E$, once stipulated, remain as stipulated throughout the measurement process.  This assumption merely reflects adherence to the PDE.  However, the usual interpretation of the von Neumann chain also assumes both that the macroscopic ready states $|A^{ready}>$ and $|O^{ready}>$ and the final pointer states $|a_{n}>$ and $|o_{n}>$ are ``given" by the oracle of decoherence, and that these ``given" pointer states refer uniquely to the stipulated $A$ and $O$.  The second of these assumptions requires the oracle of decoherence to be complete.  It therefore, by Theorem 2, requires the universe to be classical, contradicting the claim that the von Neumann chain is describing a quantum measurement.  Any interpretation of quantum measurement that assumes that the pointer states $|A^{ready}>$ and $|O^{ready}>$ and $|a_{n}>$ and $|o_{n}>$ are both given by observation and refer uniquely to the stipulated $A$ and $O$ implicitly makes such an assumption of oracular completeness, and hence describes a measurement process that could only occur in a classical universe.  Standard interpretations of quantum measurement make this assumption (e.g. \cite{schloss}); the transactional interpretation of Cramer \cite{cramer86}, in which pointer states are not given by oracle, appears to avoid it.  Any interpretation of quantum measurement that takes pointer states as ``given" by an oracle and regards macroscopic objects such as apparatus and observers as having fixed identities assumes oracular completeness, and is therefore logically inconsistent.

If the assumption of oracular completeness is dropped, Eq. 4 becomes a straightforward statement of the action of decoherence.  The ``given" macroscopic pointer states $|A^{ready}>$ and $|O^{ready}>$ at the initial observation time $t_{i}$, and $|a_{n}>$ and $|o_{n}>$ at the final observation time $t_{f}$, imply the existence of decohered systems $A|_{t_{i}}$ and $O|_{t_{i}}$ at $t_{i}$ and $A|_{t_{f}}$ and $O|_{t_{f}}$ at $t_{f}$ respectively, but no extra-theoretical assumption that these apparent systems are identical to each other or to the stipulated $A$ and $O$ is made.  Indeed as indicated by Eq. 1, the most appropriate view of these observed systems is that they are themselves quantum superpositions.  The only unique and well-defined individual entity taken to exist throughout the measurement process is the joint wave function $|S>|A>|E>|O>$, which is viewed as labeled in the way indicated by Eq. 4 strictly for convenience.  The Hamiltonian $\mathcal{H}_{\mathit{U}}$ is taken to be unitary; hence there is nothing that can be said to ``collapse" or ``branch."  In this representation, linearity is never violated, the measurement problem in its usual form does not arise, and the absence of experimental evidence of wave-function collapse \cite{schloss06} becomes positive experimental evidence supporting decompositional equivalence.

Zurek has pointed out that the measurement problem can only be formulated if a decomposition into systems is assumed \cite{zurek98rough, zurek03rev}.  As the above discussion shows, decomposition into systems is necessary, but not sufficient, to state the measurement problem in its traditional form.  The measurement problem arises only if decomposition into systems is accompanied by an assumption of oracular completeness on the part of decoherence, or on the part of whatever oracle ``gives" pointer states to observers.  Only if that oracle observationally defines unique, continuously-existing entities is there an entity to collapse or branch.  However, if decoherence or any other oracle defines unique, continuously-existing entities, the universe is classical and the notions of collapse or branching are otiose.

Interpreting measurement without an assumption of oracular completeness clearly leaves a significant question unanswered: the question of what an observer can \textit{learn} about a system $S$ by making a measurement.  This is a question about the observer's state, which to avoid an assumption of oracular completeness can only be considered as a pointer state $|o_{k}>$ at a particular observation time $t_{k}$.  All information must be physically encoded \cite{landauer91}, so $|o_{k}>$ must be regarded as encoding whatever information $O$ ``has" at $t_{k}$.  At $t_{i}$ when the measurement begins, $|o_{i}>$ must encode, among other things, the decomposition $\{S, A, E, O\}$, specifications of the pointer states $\{|a_{1}> ... |a_{N}>\}$ observable at the resolution available to $O$, a mapping from these available pointer states to computed states $\{|s_{1}> ... |s_{M}>\}$ of the stipulated $S$, and the fact that $|A^{ready}>$ is currently observed.  At $t_{f}$ when the measurement ends, $|o_{f}>$ must encode the above information, with the exception that a different pointer state $|a_{n}>$ is now indicated as currently observed.  The transition from $|o_{i}>$ to $|o_{f}>$ is implemented by $\mathcal{H}_{\mathit{U}}$, and may be investigated by examining the action of the stipulated projection $|O><O|\mathcal{H}_{\mathit{U}}$ on stipulated model observer states $|O^{ready}>$.  However, the local observer Hamiltonian $|O><O|\mathcal{H}_{\mathit{U}}$ cannot be regarded as empirically specifiable, even in principle.  To do so would introduce an assumption of oracular completeness.

The question raised by Zurek in the final paragraph of his celebrated ``Rough Guide" to decoherence \cite{zurek98rough}, that of ``how to define (systems) given, say, the overall Hamiltonian" can now be partially answered.  As Schmelzer \cite{schmelzer09} has pointed out, it is not possible to uniquely define systems that ``exist" at any time $t$ given only knowledge of the universal Hamiltonian $\mathcal{H}_{\mathit{U}}$.  However, introducing additional observables as ``fundamental" does not solve this problem; decompositional equivalence guarantees that any additional observable, if regarded as an oracle, will display the same ambiguities as the Hamiltonian.  Any addition to Hamiltonian dynamics that yields unique ``existing" systems from observations assumes oracular completeness, and hence classicality.  Systems can be arbitrarily stipulated by observers, but the dynamics of a universe compliant with decompositional equivalence, and hence allowing the definition of quantum mechanical operators, does not respect such stipulations: stipulated systems do not ``exist."  Stipulations of systems by observers do, however, represent information, and that information must be physically encoded by the states of the stipulating observers.  What systems are under observation or discussion at a given time by a given observer is, therefore, a physical fact that can be discovered from, and must be regarded as produced by, $\mathcal{H}_{\mathit{U}}$.

\section{Oracular completeness and the ``emergence of the classical"}

The decoherence program replaced the traditional notion that the classical should be a limit case of the quantum (e.g. as $\hbar \rightarrow 0$) with the somewhat more subtle notion that, under the proper observational circumstances, the classical would ``emerge" from the quantum \cite{zurek98rough, zurek03rev, schloss}.  The program of quantum Darwinism has demonstrated that multiple observers can, without prior knowledge of a system $S$, reliably expect to be able to obtain indistinguishable pointer states $|p>$ representing $S$ from distinct, disjoint fragments of the environment, even under non-optimal observational conditions \cite{zurek03rev, zurek09rev, zwolak09}.  The intuitive classical world is not, however, a world of pointer states; it is a world of ``robust" objects.  As we have seen, the only way to obtain this ``robustness" from quantum mechanics is to assume it: to assume that $S$ has a fixed boundary that the oracle of decoherence faithfully reports.  But this assumption of oracular completeness defeats everything; it is so strong that it is unwarranted even in classical physics.

The most straightforward inference from this state of affairs is that ``robust" intuitive classicality does not, after all, ``emerge" from quantum mechanics.  It is, rather, \textit{imposed} on quantum mechanics from the outside in the form of an assumption of oracular completeness.  It is reasonable to ask, therefore, where this imposed assumption comes from.  In human observers, the assumption of oracular completeness appears to derive from the assumption of object permanence, an assumption that develops early in infancy and appears to be hard-wired into the interface between human perceptual and cognitive systems \cite{ruffman05, moore-meltzoff08}.  Object permanence underlies human ``autonoetic" consciousness, which includes awareness both of episodic memories and of the continuous existence of the self as holder of those memories \cite{Sudden07, Piolino09}.  Experiments with a variety of non-human species suggest that such autonoetic consciousness is, on Earth, human-specific \cite{Sudden07, Penn08}.  Thus a hard-wired result of biological Darwinism, one expressed, as far as we can tell, only in ourselves, must be added to the pointer states produced by quantum Darwinism to get the ``robustness" of the classical world.

\section{Conclusion}

The results obtained here depend on four assumptions: 1) a universe $U$ compliant with the Principle of Decompositional Equivalence, 2) a Hamiltonian $\mathcal{H}_{\mathit{U}}$ that is everywhere well defined, 3) a finite minimum observation time $\Delta t$, and 4) a finite maximum information-transmission speed $c$.  Under these conditions, observers can stipulate the boundaries of systems of interest, but cannot determine such boundaries by observation.  Any extra-theoretical assumption that observers can uniquely specify boundaries of systems by observation is stronger than an assumption of classicality.  These results show that what requires interpretation in quantum mechanics is not the measurement process or the production of pointer states, but rather the representation by observers of ``given" observations as contributing to ``knowledge" about systems.  Such representation is a physical process resulting in a physical encoding by components of the observer state.  This encoding process is not presently well characterized; however, the understanding of the human implementation of this process by cognitive neuroscience is advancing with impressive speed.

The measurement problem as traditionally formulated is an artifact of an extra-theoretical and non-empirical assumption of oracular completeness that is inconsistent with quantum mechanics.  If this assumption is abandoned, both the measurement problem and the requirement that physics somehow produce the ``robustness" of classical objects simply dissolve.  It is ironic, given the long history of appeals to ``consciousness" to resolve the measurement problem \cite{neumann, wigner, orlov, mermin, penrose, zeh, stapp}, that consciousness, and specifically hard-wired human autonoetic consciousness, appears to be the source of our human fiction of ``robustness", and hence of our interpretational difficulties regarding measurement.

\end{document}